\documentclass[%
 reprint,
 amsmath,amssymb,
 aps,
]{revtex4-2}

\usepackage{graphicx}
\usepackage{dcolumn}
\usepackage{bm}

\begin{document}

\title{Long Spin Coherence and Relaxation Times in Nanodiamonds Milled from Polycrystalline $^{12}$C Diamond}
\author{James E March$^1$, Benjamin D Wood$^1$, Colin J Stephen$^1$, Laura Durán Fervenza$^1$, Ben G Breeze$^1$, Soumen Mandal$^2$, Andrew M Edmonds$^3$, Daniel J Twitchen$^3$, Matthew L Markham$^3$, Oliver A Williams$^2$, Gavin W Morley$^1$}

\affiliation{$^1$Department of Physics, University of Warwick, Coventry, CV4 7AL, United Kingdom}
\affiliation{$^2$School of Physics and Astronomy, Cardiff University, Queen's Building, The Parade, Cardiff, CF24 3AA, United Kingdom }
\affiliation{$^3$Element Six Global Innovation Centre, Fermi Avenue, Harwell Oxford, Didcot, Oxfordshire OX11 0QR, United Kingdom}

\date{\today}

\begin{abstract}
	The negatively charged nitrogen-vacancy centre (NV$^-$) in diamond has been utilized in a wide variety of sensing applications. The centre's long spin coherence and relaxation times ($T_2^*$, $T_2$ and $T_1$) at room temperature are crucial to this, as they often limit sensitivity. Using NV$^-$ centres in nanodiamonds allows for operations in environments inaccessible to bulk diamond, such as intracellular sensing. We report long spin coherence and relaxation times at room temperature for single NV$^-$ centres in isotopically-purified polycrystalline ball-milled nanodiamonds. Using a spin-locking pulse sequence, we observe spin coherence times, $T_2$, up 786~$\pm$~200~$\mu$s. We also measure $T_2^*$ times up to 2.06~$\pm$~0.24~$\mu$s and $T_1$ times up to 4.32~$\pm$~0.60~ms. Scanning electron microscopy and atomic force microscopy measurements show that the diamond containing the NV$^{-}$ centre with the longest $T_1$ time is smaller than 100~nm. EPR measurements give an N$_{s}$$^{0}$ concentration of 0.15 $\pm$ 0.02~ppm for the nanodiamond sample.
\end{abstract}

\maketitle
\section{Introduction}

The NV$^-$ centre has proven to be an effective tool in the fields of magnetometry \cite{RondinReview,SingleMag,EnsembleMag,SpinLockMagSensing}, thermometry \cite{SingleTemp,BioTemp,EnsembleTemp,ReviewTemp}, electrometry \cite{SingleNVElectrometry,EnsembleNVElectrometry,CCDlectrometry} and radio-frequency (RF) field sensing \cite{VectorRFDetection,SpinLockRFSensing}, amongst other sensing applications.
Key properties of the electronic spin-1 system that make it favourable for sensing include optical spin readout, optical initialisation into the $m_s = 0$ state and long spin coherence and relaxation times \cite{Doherty}.  Furthermore, as nanodiamonds containing NV$^-$ centres are biocompatible \cite{NDBiocompatibility}, they show promise for biosensing, including intracellular imaging and sensing techniques \cite{NDFluorescenceImaging,NDFreeRadicalSensing,NDRelaxometrypH,BioTemp,ND_MRI,BioApplication}. 

Of the various sensing methods that the NV$^-$ centre has been applied to, many are limited by one (or more) of the centre's spin relaxation times: the inhomogeneous relaxation time, $T_2^*$, the spin coherence time, $T_2$, and the longitudinal relaxtion time, $T_1$. For example, $T_2^*$ can limit the sensitivities of thermometry \cite{1.5usT2*micropillar} and DC magnetometry \cite{RondinReview,MagnetometrySensitivity} whereas $T_1$ limits $T_1$ relaxometry measurement sensitivity \cite{RelaxometryLongerT1,CommericalND}. $T_2$ is not only a key parameter in many sensing schemes \cite{MagnetometrySensitivity,RondinReview, NMRSensitivityT2,T2inCells}, but also in the proposed use of nanodiamonds containing a single NV$^-$ for tests of fundamental physics \cite{PRL2013,NotGavin2013, BenWoodTeeth, BoseGravity}.

The spin coherence time of the NV$^-$ centre is heavily dependent on its host material. In bulk, chemical vapour deposition (CVD), low-nitrogen diamond, $T_2>$~1~s has been reported for a single centre at 3.7~K using a dynamical decoupling scheme tailored to nearby $^{13}$C nuclear spins \cite{1secondt2}. This sample also yielded $T_1$~=~3.6~$\times~10^3$~s. At room temperature, $T_1$ can be up to 6~ms in bulk CVD diamond \cite{T1Theory,T1RoomTemp}. $T_2$~=~730~$\mu$s, obtained with a Hahn-echo sequence, has been observed for a single centre at room temperature in bulk CVD diamond with natural $^{13}$C abundance \cite{Colin2019}. This was extended to $T_2$~=~2.4~ms using dynamical decoupling \cite{Colin2019}, whereas $T_2$~=~1.8~ms has been measured for a single NV$^-$ centre using a Hahn-echo sequence in isotopically purified, ultrapure bulk CVD diamond \cite{Balasubramanian2019}. 

Spin coherence and relaxation times in nanodiamond have so far been measured to be significantly shorter. This is thought to be due to interaction with spins on the surface of the nanodiamond, as well as the typically higher nitrogen concentration found in nanodiamond \cite{KnowlesND,T1andT2Surface, T1Surface_2, T1Surface}. Surface effects are particularly detrimental to the $T_1$ of NV$^-$ centres in nanodiamonds \cite{T1Surface_2,T1Surface}. 

The longest spin coherence time observed in nanodiamond (or microdiamond) to date is 708~$\mu$s \cite{Nanograss}. This measurement was made on a single NV$^-$ centre in a single-crystal, isotopically-purified, low-nitrogen, lithographically fabricated nanopillar. The pillars were 300-500~nm in diameter and 0.5-2~$\mu$m in length. A version of these pillars (diameter = 500~nm and length = 2~$\mu$m) yielded a single NV$^-$ $T_2^*$ time of 6.42~$\mu$s \cite{Nanograss}.
$T_2$~=~ 468~$\mu$s has been demonstrated for single NV$^-$ centres in smaller, 200~nm diameter, ball-milled nanodiamonds \cite{BenWoodND}. Milling allows for the quick production of a large mass of nanodiamond as 3D volumes of bulk diamond (e.g. a mass of 0.1~g) can be processed in one go, in contrast to the 2D approach of fabricating pillars. In a different milled nanodiamond sample with a mean diameter of 23~nm, $T_2^*$ of 0.44~$\mu$s (extended to 1.27~$\mu$s using radiowaves to drive substitutional nitrogen groups) has been reported \cite{KnowlesND}. $T_1$ = 1.25~ms was measured for a single NV$^-$ centre in such a nanodiamond \cite{KnowlesND}. NV$^-$ centres in commercial nanodiamonds with diameters of around 100~nm have been found to have $T_1$ times of 100~$\mu$s or less \cite{CommericalND}.

We present long spin coherence and relaxation times for single NV$^-$ centres in small, $^{12}$C-isotopically-purified, polycrystalline, ball-milled nanodiamonds. 
We use $^{12}$C isotopically enriched diamond as our starting material to improve the NV$^-$ centres' spin coherence times. $^{13}$C, with a natural isotopic abundance in diamond of 1.1$\%$, has a nuclear spin of $1/2$ and so is a source of decoherence to the NV$^-$ centre spin \cite{Balasubramanian2019}. Using spin-locking, a pulse scheme shown to have applications including magnetometry \cite{SpinLockMagSensing}, RF field sensing \cite{SpinLockRFSensing}, paramagnetic spin bath cooling \cite{SpinBathCooling1,SpinBathCooling2} and dynamic nuclear polarisation (DNP) \cite{DNPOil,DNP2}, we observe spin-coherence times, $T_2$, up to 786~$\pm$~200~$\mu$s. Furthermore, we report $T_1$ values up to 4.32~$\pm$~0.60~ms and free-induction decay $T_2^*$ values up to 2.06~$\pm$~0.24~$\mu$s. 

\section{Methods}

The nanodiamonds used in this study were ball-milled with silicon nitride, following the process described in \cite{BallMillingCardiff}. The starting material was pieces from the edge of a CVD plate grown by Element Six. The plate was a single-crystal sample, grown for NV$^-$ magnetometry. The material from the plate's edges, however, was polycrystalline. The sample was not irradiated or annealed.

Measurements on the single NV$^-$ centres were carried out on a home-built room-temperature confocal fluorescence microscope (CFM) running QUDI software \cite{QUDI}. The nanodiamonds were deposited onto an n-type silicon wafer with a coordinate system etched into the surface. The coordinate system allowed a specific nanodiamond, investigated on the CFM, to be located using scanning electron microscopy (SEM), in order to measure its dimensions. To deposit the nanodiamonds onto the surface of the wafer, a suspension of nanodiamonds in methanol was sonicated for thirty minutes before being sprayed into a vial through a nebuliser. The vial, now containing a cloud of nanodiamonds, was upturned over the wafer, allowing the nanodiamonds to fall onto the surface of the silicon. This method was found to give a reasonably even coverage of nanodiamonds across the surface of the wafer. 

Microwaves, generated with a Keysight N5172B, were delivered to the NV$^-$ centre via a 20~$\mu$m diameter wire, brought within roughly 20~$\mu$m of the respective nanodiamond. Only nanodiamonds containing a single NV$^-$ centre were investigated. NV$^-$ centres were confirmed to be a single centre via Hanbury Brown and Twiss (HBT) experiments, where the second-order correlation function, $g^{(2)}(0)$, was below 0.5 in each case \cite{HBT0.5}.  

A static magnetic field was applied to the centre via a neodymium magnet, positioned such that the direction of the magnetic field was aligned with the nitrogen-vacancy axis. Mounting the magnet on a servo-controlled robotic arm allowed for precise control of the position and orientation of the magnet with respect to the nanodiamond. The field was chosen so that the energy gap between the $m_s$~=~0 and $m_s$~=~$-1$ sub-levels was approximately 2~GHz. This frequency was measured using continuous wave ODMR and corresponds to a magnetic field component parallel to the N-V axis, $B_\parallel$, of 62~mT at the NV$^-$ centre.

In a sample where 54 fluorescent sites were examined on the confocal microscope, 16 were confirmed to be single NV$^-$ centres via HBT experiments. As single NV$^-$ centres were the subject of this study, there was some selection bias in avoiding particularly bright spots that were likely to contain large numbers of NV$^-$ centres. It is not known how many nanodiamonds contained zero NV$^-$ centres as they do not fluoresce. The centres showed high levels of charge stability, with very few displaying charge state switching. As this was a nanodiamond sample, the N-V axis can lie in any orientation. However, the robotic arm used for aligning the magnet to the N-V axis has a relatively limited range of motion. Therefore, the specific NV$^-$ centres selected for spin coherence and relaxation time measurements were chosen because their N-V axis was in an orientation accessible to our magnetic field.

\begin{figure*}
\includegraphics[width=\linewidth]{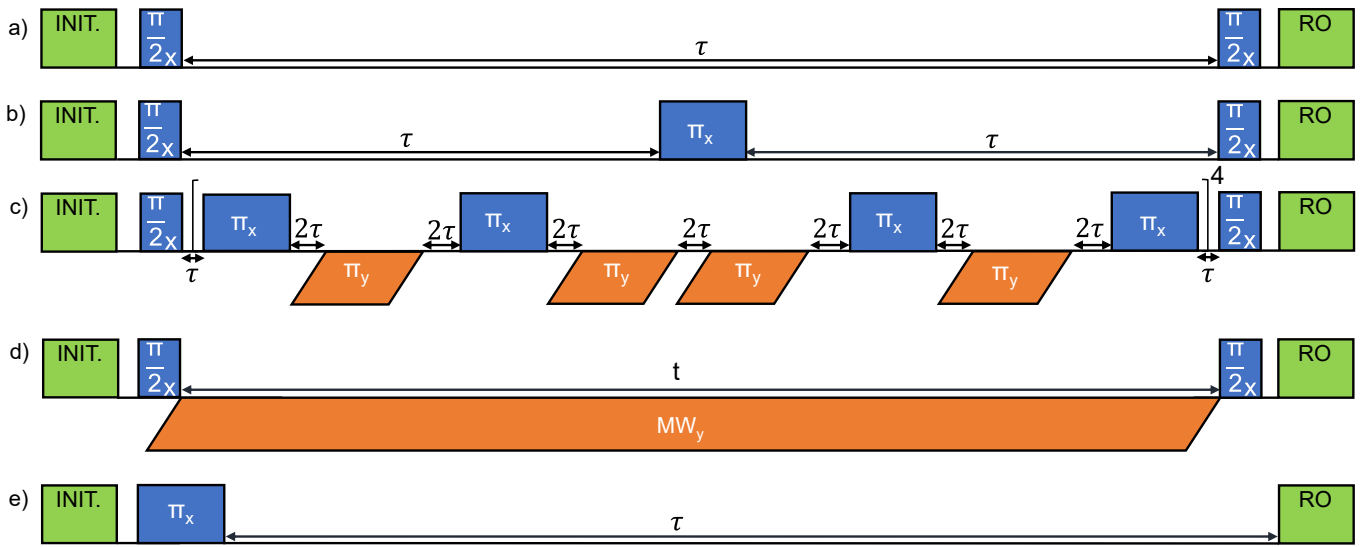}
\caption{\label{fig:PulseSequences} Schematic of the five different pulse sequences used in this study. The green block at each end of each sequence represents the 532~nm laser initialisation (INIT.) and readout (RO) pulses, which are identical for each pulse sequence. In practice, the initialisation and read out is performed by the same (single) pulse, as the pulse sequence is repeated many times to build up the signal-to-noise ratio. The blue and orange blocks represent x and y-microwave pulses respectively. a) shows the Ramsey sequence, used to measure $T_2^*$. b) shows the Hahn-echo sequence, used to measure $T_2$. c) shows an XY8-4 pulse sequence. The section of eight $\pi$-pulses repeats four times before the second $\pi/2$ pulse is applied. This sequence was used to measure $T_2^{(XY8-4)}$. d) shows the spin-locking pulse sequence, used to measure $T_2^{(SL)}$. e) shows the sequence used to measure $T_1$. Phase cycling was used in sequences a - d to reject common-mode noise.}
\end{figure*}

Five NV$^-$ centres in five different nanodiamonds were fully characterised in this study. $T_2^*$ was measured for each nanodiamond, as well as three measurements of $T_2$ using Hahn-echo, XY8-4 and spin-locking pulse sequences. $T_1$ was also measured for each NV$^-$. The pulse sequences used to perform the measurements are displayed in Figure~\ref{fig:PulseSequences}.

All of the pulse sequences displayed in Figure~\ref{fig:PulseSequences} begin and end with a 532~nm laser pulse. These are the initialisation (into the $m_s$~=~0 state) and readout pulses respectively. Figure~\ref{fig:PulseSequences}a shows a Ramsey scheme, used to measure the inhomogeneous dephasing time, $T_2^*$, of the NV$^-$ centre. The  $T_2^*$ of a single NV$^-$ centre is primarily governed by slowly-varying magnetic and electric fields as well as strain and temperature fluctuations \cite{MagnetometrySensitivity}. The Hahn-echo sequence, Figure~\ref{fig:PulseSequences}b, largely negates these dephasing mechanisms. The $\pi$ pulse inverts the spin's precession, and so phase accumulated during the first wait period $\tau$ is cancelled in the second \cite{MagnetometrySensitivity}. This is provided that the timescale on which the noise varies is long compared to $\tau$. The resulting decay constant, $T_2$, is dominated by magnetic interaction with proximate spins.

In Figure~\ref{fig:PulseSequences}c, the XY8 sequence, additional refocusing $\pi$ pulses mitigate the dephasing effects of magnetic noise fluctuating on shorter timescales \cite{CPMG}. The phase switching and time symmetry of the sequence means that the scheme is more resistant to errors in pulse length than, for example, a CPMG (Carr-Purcell-Meiboom-Gill) sequence \cite{XY8vsCPMG,RobustnessofDDschemes}. 

Figure~\ref{fig:PulseSequences}d shows a spin-locking pulse sequence. Also referred to as $T_{1\rho}$ ($T_1$ in the rotating frame), it is considered to be the upper limit of $T_2$ that can be measured by a dynamical decoupling sequence, for a given microwave power \cite{NaydenovSL}\cite{SurfaceSpins3}. After initialisation into the $m_s$~=~0 state, a $(\pi/2)_x$ pulse places the spin of the NV$^-$ centre into an equal superposition of $m_s$ = 0 and $m_s$ = 1. A long microwave pulse, phase-shifted by 90$^{\circ}$ to the $(\pi/2)_x$ pulse, continuously drives the spin, keeping it aligned along the y-axis of the Bloch sphere. The spin-state is optically read out following a further $\pi$/2 pulse. Figure~\ref{fig:PulseSequences}e shows the $T_1$ sequence.  

\begin{figure*}
\includegraphics[width=\linewidth]{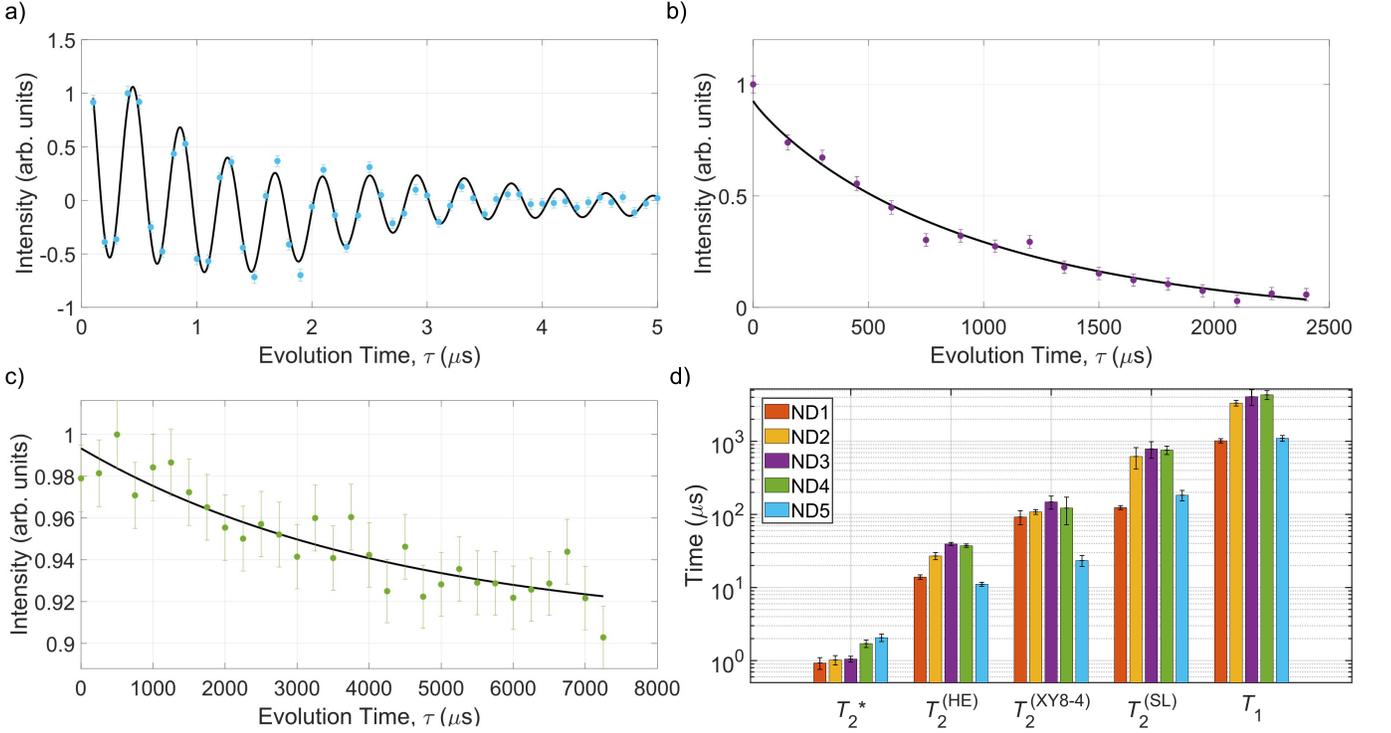}
\caption{\label{fig:DataClumped} a)  Ramsey measurement of $T_2^*$ coherence time for ND5 using the pulse sequence shown in Figure~\ref{fig:PulseSequences}a. $T_2^* = 2.06\pm0.24~\mu$s was extracted from the fit of the form $A~+~Be^{-(\tau/T_2^*)}\Sigma_ksin(c_kt + d_k)$ with three frequency components \cite{T2*Fit}. b) Spin-lock measurement of $T_2^{(SL)}$ coherence time for ND3 using the pulse sequence shown in Figure~\ref{fig:PulseSequences}d. A value of $T_2^{(SL)}~=~786\pm200~\mu$s was extracted from a fit of the form $A~+~Be^{-(\tau/T_2)}$ \cite{SpinLockRFSensing}. c) Measurement of $T_1$ relaxation time for ND4 using the pulse sequence shown in Figure~\ref{fig:PulseSequences}e. $T_1$~=~$4.32\pm0.60$~ms was extracted from the exponential fit of the form $A~+~Be^{-(\tau/T_1)}$ \cite{T1Fit}. d) Plot showing the measured values of $T_2^*$, $T_2$ (obtained through Hahn-echo, XY8-4 and spin-locking pulse sequences) and $T_1$ for each of the five NV$^-$ centres investigated in this study. The mean value for each of the five measured quantities across the five NV$^-$ centres is as follows: $T_2^*$~=~1.35~$\pm$~0.20~$\mu$s, $T_2^{(HE)}$~=~25.8~$\pm$~5.2~$\mu$s, $T_2^{(XY8-4)}$~=~98.7~$\pm$~19~$\mu$s, $T_2^{(SL)}$~=~494~$\pm$~96~$\mu$s and $T_1$~=~2.77~$\pm$~0.64~ms.} 
\end{figure*}

\section{Results}

Of the five centres in this study, the longest recorded $T_2^*$, $T_2$ and $T_1$ measurements are displayed in Figure~\ref{fig:DataClumped}a-c respectively. The longest measured $T_2$ was obtained via the spin-locking pulse sequence. Figure~\ref{fig:DataClumped}d shows $T_2^*$, $T_2^{(HE)}$, $T_2^{(XY8-4)}$, $T_2^{(SL)}$ and $T_1$ measurements for each of the five NV$^-$ centres. We find that the same NV$^-$ centre (ND3) yields the longest times for $T_2^{(HE)}$, $T_2^{(XY8-4)}$ and $T_2^{(SL)}$, whereas the longest recorded $T_2^*$ and $T_1$ were obtained from two other centres (ND5 and ND4 respectively). It was found that spin-locking can increase $T_2$ by up to a factor of 5.7 relative to XY8-4. A sixth single NV$^-$ was partially characterised and was found to have $T_2^{(HE)}$~=~12.4~$\pm$~0.5~$\mu$s, comparable to ND1, ND2 and ND3. However, due to a combination of low centre fluorescence, low ODMR contrast and sample drift, it was not possible to make dynamical decoupling measurements at the same microwave power that had been used in measurements on the other five centres, or to measure its $T_1$. No spin measurements were made for any other NV$^-$ centres from this nanodiamond sample.

\begin{figure*}[!htbp]
\includegraphics[width=\linewidth]{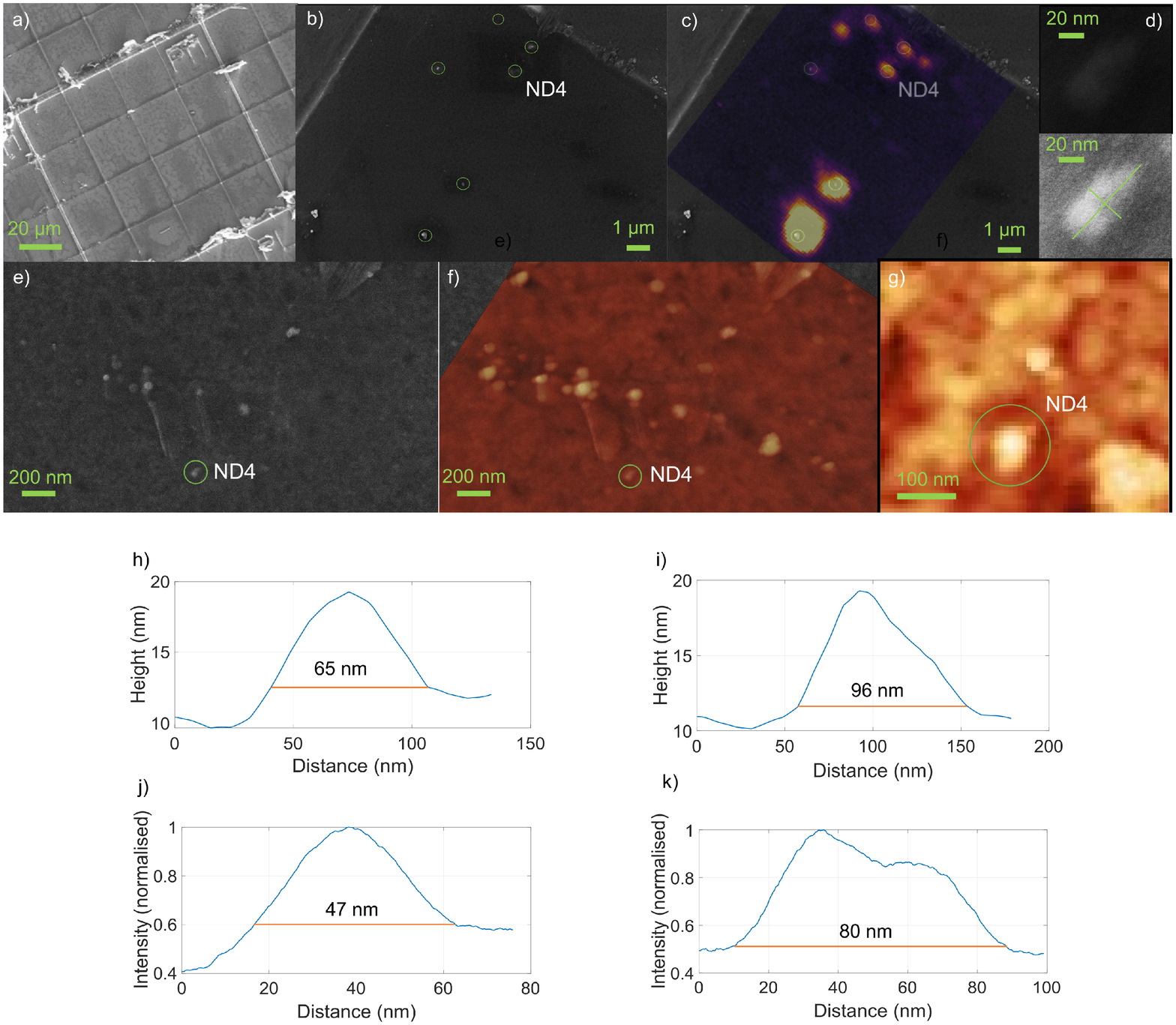}
\caption{\label{fig:SEM} a) SEM image showing the coordinate system on the surface of the silicon wafer. The system allows specific nanodiamonds, known to contain single NV$^-$ centres from examination on a CFM, to be located using SEM. b) SEM image showing the location of ND4. The NV$^-$ centre contained within the highlighted nanodiamond (ND4) had the following spin coherence and relaxation times: $T_2^*$~=~1.71~$\pm$~0.20~$\mu$s, $T_2^{(SL)}$~=~759~$\pm~100~\mu$s and $T_1$~=~4.32~$\pm~0.60~$ms. c) A fluorescence image, taken on a CFM, of the same area as b), overlaid on b). This shows how nanodiamonds containing fluorescent NV$^-$ centres can be located using both SEM and CFM. Filters (532~nm notch and 633~nm high-pass) are used on the CFM to optimise for collection of light emitted from the NV$^-$ centre, which has an optical zero-phonon line at 637~nm \cite{Doherty}.
d) High-magnfication images of ND4. The upper image shows the original SEM image, whereas the lower image has been enhanced in ImageJ to allow for easier visualistion of its dimensions. The lower image was enhanced by using the `equalize histrogram' function and then applying a Gaussian blur with $\sigma$ = 1.5). 
e) High-magnification SEM image showing the location of ND4. f) AFM image of the area shown in e), overlaid over e), showing the agreement between AFM and SEM images of the area. g) High-magnification AFM image of ND4. h)  Plot of height against position for the short axis of the diamond in the x-y plane for the AFM image shown in g). i) Plot of height against position for the long axis of the diamond in the x-y plane for the AFM image shown in g). The surface of the silicon is at a height of 11.5~$\pm$~1.1~nm, giving ND4 a height of 7.8~$\pm$~1.1~nm. j) Plot of pixel intensity against position for the short axis of the diamond in the x-y plane for an SEM image (shown in d)). The length of the short axis is $47~\pm$~10~nm. k) Plot of pixel intensity against position for the long axis of the diamond in the x-y plane for an SEM image (as shown in d)). The length of the long axis is $80~\pm$~10~nm. The measurements in the x-y plane taken from the SEM images are shorter than those taken from the AFM images, as the AFM images are the convolution of the shape of the diamond with the shape of the AFM probe.} 
\end{figure*}

Due to the aforementioned coordinate system on which the diamonds were deposited, it was possible to locate and measure the dimensions of ND4 ($T_2^*$~=~1.71~$\pm$~0.20~$\mu$s, $T_2^{(SL)}$~=~759~$\pm~100~\mu$s and $T_1$~=~4.32~$\pm~0.60~$ms) using atomic force microscopy (AFM) and SEM. The long and short axes of the diamond in the x-y plane were measured to be $80~\pm~10$~nm and $47~\pm~10$~nm respectively, as shown in Figure~\ref{fig:SEM}. 
The height of ND4 was measured, using AFM, to be 7.8~$\pm$~1.1~nm. Therefore, the maximum distance that the NV$^-$ centre can be from a surface of the diamond is less than 5~nm. This is comparable to the dimensions of the diamonds investigated in \cite{KnowlesND} and far smaller than those studied in \cite{Nanograss}. 
Unfortunately, due to a combination of surface contamination and nanodiamond agglomeration, it wasn't possible to locate any of the other four diamonds referenced in Figure \ref{fig:DataClumped}. 

\section{Discussion}

To our knowledge, the values of $T_2^*=2.06~\pm~0.24~\mu$s, $T_2 = 786~\pm~200~\mu$s and $T_1 = 4.32\pm0.6~$ms are the longest measurements of these three characteristics for an NV${^-}$ centre in nanodiamond. Furthermore, we believe that $786~\pm~200~\mu$s is the longest $T_2$ reported for any electronic spin system in a nanoparticle. Although  $T_2^* = 6.42~\pm~1.05~\mu$s has been reported for a single NV$^-$ in a high-purity fabricated diamond pillar, the diamond was microscale as opposed to nanoscale \cite{Nanograss}. Many sensing applications for NV$^-$ centres in nanodiamond require nanodiamonds with diameters of around 100~nm or less, for example, those where the diamonds enter cells \cite{NDUnder100nmBiocompatible}. As our nanodiamonds were milled rather than etched, we efficiently created large quantities of nanodiamond, on the order of 0.1~g. On a similar note, polycrystalline CVD diamond is easier to grow than single crystal, as well as being cheaper.

The average $T_2^*$ (1.35~$\mu$s) across the five NV$^-$ centres included in this study was longer than the average ensemble $T_2^*$ (1.13~$\mu$s) measured for a bulk sample of isotopically-purified, electron-irradiated-and-annealed, polycrystalline CVD diamond with a subsititutional nitrogen impurity concentration of 16~ppm \cite{EdmondsPoly}. 
Using EPR, the N$_{s}$$^{0}$ concentration of our sample was measured to be 0.15 $\pm$ 0.02~ppm. This lower nitrogen concentration may explain why the NV$^-$ centres in these nanodiamonds have a longer average $T_2^*$ time than NV$^-$ centres in the bulk sample characterised in \cite{EdmondsPoly}. For this isotopically purified nanodiamond sample, substitutional nitrogen impurities give rise to dephasing due to intrinsic electronic spins \cite{NitrogenDominantDephasing, MagnetometrySensitivity}. There may also be a significant dephasing contribution from magnetic noise arising from effects at the diamond's surface \cite{SurfaceSpins1,SurfaceSpins2,KnowlesND,SurfaceSpins3,DeLeon}. It is not known which of these is the dominant dephasing mechanism. In the event that it is substitutional nitrogen, we would expect to see an increase in spin coherence times by using a starting material with a lower nitrogen concentration, and then irradiating and annealing to create a similar NV$^-$ concentration \cite{IrradiateAndAnneal, NConcentration}.

As NV$^-$ sensing techniques are often limited by one or more of $T_2^*$, $T_2$ and $T_1$, as well as how close the NV$^-$ centre can be brought to the target, making these times long for NV$^-$ centres in nanodiamond is favourable for many sensing applications \cite{MagnetometrySensitivity,NVRelaxometryReview,NDInCellsReview_2}. Furthermore, cellular uptake and biocompatibility has been demonstrated for nanodiamonds similar in size to ND4 \cite{CellularUptakeND,NDInCellsReview}, and so such diamonds are favourable for intracellular sensing techniques. An example is NV$^-$ $T_1$ relaxometry, which has shown promise for detecting free radicals and measuring pH changes in cells, amongst other things \cite{FreeRadicalDetection,FreeRadicalDetection_2, pHDetection_2, pHDetection, NVRelaxometryReview,NDBiosensingReveiw}. As the sensitivity is limited by the $T_1$ and proximity to the diamond surface of the NV$^-$ centres, the availability of smaller diamonds hosting NV$^-$ centres with longer $T_1$ offers new opportunities \cite{NVRelaxometryReview, NDInCellsReview_2}. The shortest $T_1$ measured in this study ($1.1~\pm$~0.1~ms) is comparable to the longest $T_1$ currently reported in the literature for a single NV$^-$ centre in nanodiamond (1.25~ms)\cite{KnowlesND}. 

\begin{figure*}[htb]
\includegraphics[width=\linewidth]{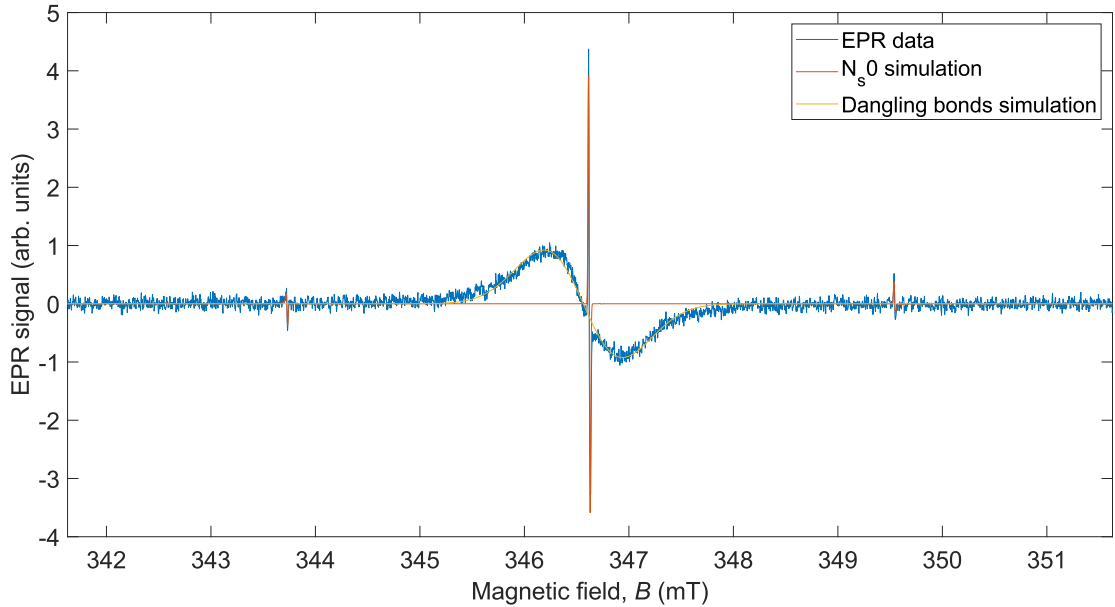}
\caption{\label{fig:EPRData} EPR data for a 40~mg nanodiamond sample. The blue line shows the actual EPR data. The red and yellow lines show simulated EPR signals resulting from N$_{s}^{0}$ and dangling bonds respectively. The simulations were done using Easyspin \cite{Easyspin}. We find concentrations of 150~$\pm$~15~ppb and 500~$\pm$~50~ppb for N$_{s}^{0}$ and dangling bonds respectively.}
\end{figure*}

\section{Conclusion}
We have presented long spin relaxation and coherence times for single NV$^-$ centres hosted in nanodiamonds. The nanodiamonds were ball-milled with silicon nitride from polycrystalline, isotopically purified CVD diamond with an N$_{s}$$^{0}$ concentration of 0.15 $\pm$ 0.02~ppm. The long axis, short axis and height were measured to be $80\pm10$~nm, $47\pm10$~nm and $7.8~\pm~1.1$~nm respectively for a particular diamond hosting an NV$^-$ centre known to have $T_2^*$~=~1.71~$\pm$~0.20~$\mu$s, $T_2^{(SL)}$~=~759~$\pm~100~\mu$s and $T_1$~=~4.32~$\pm~0.60~$ms. Small diamonds containing NV$^-$ centres with good spin coherence properties, that can be produced in large quantities, are favourable for a wide variety of sensing applications \cite{NVRelaxometryReview,NDInCellsReview_2,NDRelaxometrypH,FreeRadicalDetection,NDElectricFields,NDIonSensing}.

\section{Acknowledgements}
We would like to thank Ben Green and Steve York for useful discussions. J. E. M's studentship is supported by the Royal Society. B. D. W's PhD studentship is supported by the Engineering and Physical Sciences Research Council (EPSRC). G. W. M is supported by the Royal Society. The work is supported by the following grants from the UKRI EPSRC: EP/M013243/1 (UK National Quantum Technologies Programme, NQIT Hub), EP/T001062/1 (QCS Hub),  EP/M013294/1 (Quantum Technology Hub for Sensors and Metrology), EP/V056778/1 (Prosperity Partnership) and EP/L015315/1 (EPSRC CDT in Diamond Science and Technology), EP/V007688/1 (Warwick Analytical Science Centre) . The work is supported by a Science and Technologies Facilities Council (STFC) grant with grant number 
ST/W006561/1.

\appendix*
\section{}

Figure \ref{fig:EPRData} displays data from EPR measurements taken on the nanodiamond sample. The measurements were made on a 40~mg nanodiamond sample using a Bruker EMX spectrometer with a Bruker SHQ-E cavity. The modulation frequency and amplitude were 100~kHz and 0.01~mT respectively. Easyspin was used for the data processing and simulations \cite{Easyspin}. The EPR data was compared against a well-characterised Type 1b diamond and the nanodiamond sample used in this study was found to have an N$_{s}^{0}$ concentration of 150~$\pm$~15~ppb. The EPR signal has a linewidth of 0.011~mT. We attribute the broad features visible in the spectrum to dangling bonds on the nanodiamond surface, for which we measure a concentration of 500~$\pm$~50~ppb \cite{DB1,DB2,DB3, DB4}.

\providecommand{\noopsort}[1]{}\providecommand{\singleletter}[1]{#1}
%



\end{document}